\documentclass[conference]{IEEEtran}

\usepackage{setspace}
\usepackage{cite}
\usepackage{citesort}
\usepackage{graphicx}
\usepackage{psfrag}
\usepackage{subfigure}
\usepackage{url}
\usepackage{stfloats}
\usepackage{amsmath}
\usepackage{amsfonts}
\usepackage{amssymb}
\usepackage{array}

\begin{document}

\title{\huge Performance of Spatial Modulation using Measured Real-World Channels\vspace{-0.1cm} }

\author{\authorblockN{A. Younis$^{(1)}$, W. Thompson$^{(2)}$, M. Di Renzo$^{(3)}$,  C.-X.~Wang$^{(4)}$, M. A. Beach$^{(2)}$, H. Haas$^{(1)}$, and P. M. Grant$^{(1)}$}
 \IEEEauthorblockA{$^{(1)}$\normalsize\sl Institute for Digital Communications, The University of Edinburgh  \\ King's Buildings, Mayfield Road, Edinburgh. EH9 3JL, UK }
 \IEEEauthorblockA{$^{(2)}$\normalsize\sl
 The University of Bristol, Bristol, BS8 1UB, UK}
 \IEEEauthorblockA{$^{(3)}$\normalsize\sl 
 \'Ecole Sup\'erieure d'\'Electricit\'e (SUP\'ELEC), University of Paris--Sud XI (UPS)\\
 3 rue Joliot--Curie, 91192 Gif--sur--Yvette (Paris), France}
 \IEEEauthorblockA{$^{(4)}$\normalsize\sl  School of Engineering \& Physical Sciences \\ Heriot-Watt University, Edinburgh, EH14 4AS, UK \\
 E--Mail: \{a.younis, h.haas, p.grant\}@ed.ac.uk, \{w.~thompson, m.a.beach\}@bristol.ac.uk,\\ marco.direnzo@lss.supelec.fr, cheng-xiang.wang@hw.ac.uk}  \vspace{-0.9cm} }
 
\maketitle

\begin{abstract}

In this paper, for the first time real-world channel measurements are used to analyse the performance of spatial modulation (SM), where a full analysis of the average bit error rate performance (ABER) of SM using measured urban correlated and uncorrelated Rayleigh fading channels is provided. The channel measurements are taken from an outdoor urban multiple input multiple output (MIMO) measurement campaign. Moreover, ABER performance results using simulated Rayleigh fading channels are provided and compared with a derived analytical bound for the ABER of SM, and the ABER results for SM using the measured urban channels. The ABER results using the measured urban channels validate the derived analytical bound and the ABER results using the simulated channels.
 Finally, the ABER of SM is compared with the performance of spatial multiplexing (SMX) using the measured urban channels for small and large scale MIMO. It is shown that SM offers nearly the same or a slightly better performance than SMX for small scale MIMO. However, SM offers large reduction in ABER for large scale MIMO.
 
\end{abstract}

 \vspace{-0.4cm}

\IEEEpeerreviewmaketitle
\section{Introduction} \label{sec1}
 
 Multiple input multiple output (MIMO) systems offer a significant increase in spectral efficiency in comparison to single antenna systems \cite{t9902}. An example is spatial multiplexing (SMX) \cite{f9601}. SMX achieves a spectral efficiency that increases linearly with the number of transmit antennas, by transmitting simultaneously over all the transmit antennas. However, SMX  cannot cope with the exponential increase of wireless data traffic, and a larger number of transmit antennas (large scale  MIMO) should be used.
  Large scale MIMO systems studied in \cite{cv0901}, offers a higher data rate and better average bit error performance (ABER). However, this comes at the expense of an increase in i) hardware complexity, where the number of radio frequency (RF) chains is equal to the number of transmit antennas, ii) receiver computational complexity, where the SMX complexity increases exponentially with the number of transmit antennas. Thus, SMX may not be always feasible and a cheaper solution should be used.

 Spatial Modulation (SM) is a transmission technology proposed for MIMO wireless systems. It aims to increase the spectral efficiency, of single--antenna systems while avoiding Inter--Channel Interference (ICI) \cite{mhsay0801}. This is achieved by i) the activation of a single antenna at each time instance which transmits a given data symbol (\emph{constellation symbol}), and ii) the exploitation of the spatial position (index) of the active antenna as an additional dimension for data transmission (\emph{spatial symbol}). The receiver applies the Maximum Likelihood optimum decoder for SM (SM--ML), which performs an exhaustive search over the whole \emph{constellation symbol} and \emph{spatial symbol} space. Activating only one antenna at a time means that only one RF chain is needed, which significantly reduces the hardware complexity of the system \cite{jgsc0901}. It also offers a significant reduction in the energy needed. This reduction increases linearly with the number of transmit antennas, as only one antenna needs to be powered at a time, \emph{i.e.}, ``green'' technology. Moreover, as it will be shown in this paper the computational complexity of SM--ML is equal to the complexity of single--antenna systems, \emph{i.e.}, the complexity of SM--ML does not depend on the number of transmit antennas. Accordingly, SM is an attractive candidate for large scale  MIMO.

 In this paper, for the first time real-world channel measurements are used to analyse the performance of SM, where a full analysis of the ABER of SM using measured urban correlated and uncorrelated Rayleigh fading channels is provided. The channel measurements are taken from an outdoor urban MIMO measurement campaign. Moreover, an analytical bound for the ABER of SM is derived and performance results using simulated Rayleigh fading channels are provided. It is shown that the results using the measured urban channels validate the derived analytical bound and the results using the simulated channels.
 Furthermore, the ABER of SM is compared with the performance of SMX using the measured urban channels for small and large scale MIMO. It is shown that SM offers nearly the same or a slightly better performance than SMX for small scale MIMO. However, SM offers large reduction in ABER for large scale MIMO.

 The remainder of this paper is organised as follows. In Section \ref{SM}, the system model and the ML--optimum receiver are summarised. In Section \ref{Cha_Mod}, the channel measurements are introduced. In Section \ref{ANA}, an analytical bound for SM over correlated and uncorrelated Rayleigh channels is derived. The complexity of SM and SMX is discussed and compared in \ref{comp}. Finally, the results are presented in Section \ref{results}, and the paper is concluded in Section \ref{Con}.

\section{System Model} \label{SM}

 In SM, the bit stream emitted by a binary source is divided into blocks containing $m=\log_2 \left( {N_t } \right)
+ \log_2 \left( M \right)$ bits each, where $m$ is the spectral efficiency, $N_t$ is the number of transmit antennas and $M$ is the signal constellation size.
Then the following mapping rule is used \cite{mhsay0801}:
\begin{itemize}
\item The first $\log_2\left(N_t\right)$ bits are used to select the antenna that is switched on for data transmission, while all the other transmit--antennas are kept silent. In this paper, the actual transmit--antenna that is active for transmission is denoted by $\ell_t$, with
$\ell_t \in \left\{1, 2, \ldots, N_t \right\}$.
\item The second $\log_2\left(M\right)$ bits are used to choose a symbol in the signal--constellation diagram. Without loss of generality, Quadrature Amplitude Modulation (QAM) is considered. In this paper, the actual complex symbol emitted by the transmit--antenna $\ell_t$ is denoted by $s_t$, with $s_t \in \left\{ s_1, s_2, \ldots, s_M \right\}$.
\end{itemize}
Accordingly, the $N_t \times 1$ dimensional transmit vector is:
\begin{equation}
\label{Eq_1} {\bf{x}}_{\ell_t ,s_t }  = \left[ {{\bf{0}}_{1 \times \left( {\ell_t  - 1} \right)} ,s_t ,{\bf{0}}_{1 \times \left( {N_t  - \ell_t }
\right)} } \right]^T
\end{equation}
\noindent where $\left[  \cdot  \right]^T$ denotes transpose operation, and ${\bf{0}}_{p \times q}$ is a $p \times q$ matrix with all--zero entries.

The transmitted vector, ${\bf{x}}_{\ell_t ,s_t }$, in (\ref{Eq_1}) is transmitted over a flat fading $N_r\times N_t$ MIMO  channel with transfer function $\mathbf{H}$, where $N_r$ is the number of receive antennas. The Kronecker channel model \cite{kspmf0201}, with an exponential correlation profile for both the transmitter correlation matrix $(\mathbf{R}_{\text{Tx}})$ and receiver correlation matrix $(\mathbf{R}_{\text{Rx}})$, is used to model channel correlation \cite{vzh0201},
 \begin{equation}
  \mathbf{H} = \mathbf{R}_{\text{Rx}}^{\frac{1}{2}} \bar{\mathbf{H}} \mathbf{R}_{\text{Tx}}^{\frac{1}{2}}
 \end{equation}
where $\bar{\mathbf{H}}$  has independent and identically distributed (i.i.d.) entries according to $\mathcal{CN}(0, 1)$.

Thus, the $N_r \times 1$ dimensional receive vector can be written as follows:
\begin{equation}
\label{Eq_2} {\bf{y}} = {\bf{Hx}}_{\ell_t ,s_t }  + {\bf{n }}
\end{equation}
\noindent where ${\bf{n }}$ is the $N_r$--dimensional Additive White Gaussian Noise (AWGN) with zero--mean and variance $\sigma^2$ per dimension at the receiver input.

 At the receiver the ML optimum detector for MIMO systems is used,
\begin{equation}
 \hat{\mathbf{x}}_t =  \mathop {\arg \min }\limits_{\scriptstyle \mathbf{x} \in \mathcal{Q}^m} \left\{ {\left\| {{\bf{y}} - {\bf{H}}\mathbf{x}} \right\|_{\rm{F}}^2 } \right\} \label{Eq:MLMIMO}
\end{equation}
where $\mathcal{Q}^m$ is a $2^m$ space containing all possible $\left(N_t\times 1\right)$ transmitted vectors, $\left\|  \cdot  \right\|_{\rm{F}}$ is the Frobenius norm, and ${\hat  \cdot }$ denotes the estimated spatial and
constellation symbols.

In SM only one transmit antenna is active at a time. Therefore, the optimal receiver in \eqref{Eq:ML} can be simplified to,
\begin{equation}
\begin{split}
 \left[ {\hat \ell_t ,\hat s_t } \right] &= \mathop {\arg \min }\limits_{\scriptstyle \ell \in \left\{ {1,2, \ldots N_t } \right\} \hfill \atop
  \scriptstyle s \in \left\{ {s_1 ,s_2 , \ldots s_M } \right\} \hfill} \left\{ {\left\| {{\bf{y}} - {\bf{h}}_\ell s} \right\|_{\rm{F}}^2 } \right\} \\
  &= \mathop {\arg \min }\limits_{\scriptstyle \ell \in \left\{ {1,2, \ldots N_t } \right\} \hfill \atop
  \scriptstyle s \in \left\{ {s_1 ,s_2 , \ldots s_M } \right\} \hfill} \left\{ {\sum\limits_{r = 1}^{N_r } {\left| {y_r  - h_{\ell,r} s} \right|^2 } } \right\} \label{Eq:ML}\\
 \end{split}
\end{equation}
\noindent where $y_r$ and $h_{\ell,r}$ are the $r$--th entries of ${\bf{y}}$ and ${\bf{h}}_{\ell}$ respectively.

\section{Channel Measurement and Model} \label{Cha_Mod}

 The channel measurements used within this paper were acquired within the Mobile VCE MIMO elective \cite{hb0601}. MIMO channel measurements were taken around the centre of Bristol in the United Kingdom, using a MEDAV RUSK channel sounder, a $4\times4$ antenna configuration, with $20$~MHz bandwidth centred at $2$ GHz. The transmitter consisted of a pair of dual polarised ($\pm 45 ^\circ$) Racal Xp651772  antennas \cite{l1201} separated by $2$ m, positioned atop a building, providing elevated coverage of the central business and commercial districts of Bristol city. At the receiver two different receiver devices are used, both equipped with four antennas.

 The two receiver devices are a reference headset and a laptop. The reference antenna design is based on $4$-dipoles mounted on a cycle helmet, thus avoiding any shadowing by the user. The laptop is equipped with $4$ PIFA elements, both devices are detailed in \cite{hb0601}. Fifty--eight measurement locations were chosen around the city. At each location the user walked, holding the laptop in front of them and the reference device on their head, in a straight line roughly $6$ m long, until $4096$ channel snapshots have been recorded. A second measurement was then taken with the user walking a second path perpendicular to the first. As the measurement speed is significantly faster than the coherence time of the channel, the measurements are averaged in groups of four to reduce measurement noise.

 One set of measurement results with the laptop and reference device, and a second set of only the reference device measurements taken at the same locations, but on different days, is also included in the measurement data for analysis. This provides a total of $348$ different measurement sets, each containing $1024$ snapshots of a $4\times4$ MIMO channel, with $128$ frequency bins spanning the $20$~MHz bandwidth. As the simulations are carried out using flat fading channels, a single frequency bin centred around $2$~GHz, is chosen from each measurement snapshot to create the narrowband channel.

\subsection{Small Scale MIMO}
  For small scale MIMO, Rayleigh fading channels were distinguished using the Chi-squared goodness of fit test, with a significance level of $1\%$, where of the $348$ measurements, only $20$ measurements fulfilled this requirement. For each measurement the transmit and receive correlation matrices are estimated, then the decay of the correlation, based on the antenna indices, is fitted to an exponential decay model~\cite{vzh0201},
 \begin{equation} \label{eq6:CorrMod}
 \mathbf{R}_{c}=\left[
\begin{array}{cccccc}
1 & r_c & r_c^2  & \cdots & r_c^{n-1} \\	
r_c & 1 & r_c & \ddots & \vdots \\	
\vdots & \ddots & \ddots & \ddots & r_c \\	
r_c^{n-1} & \cdots & r_c^2 & r_c & 1 \\
\end{array} \right]
\end{equation} 
\noindent where $r_c = \exp\left(-\beta\right)$, $\beta$ is the correlation decay coefficient, and $n$ is the number of antennas. Two channels with the lowest mean square error between the exponential decay in \eqref{eq6:CorrMod} and the actual correlation matrices are chosen for the two correlated channel results. Both of the chosen channels are from measurements taken using the laptop device, and the measured decay coefficients for the transmitter and receiver are $0.5$ and $0.8$ for the first channel and $0.7$ and $0.4$ for the second channel respectively. 

 For the uncorrelated channels, the two channels with the lowest average correlation coefficient between their MIMO channels were chosen. One is from the laptop measurements, and the other from the reference headset device measurements. Selecting the channels in this manner may not provide completely uncorrelated channels per say, as there may still be a small correlation between the channels. However, it will provide the channel measurements that experienced the lowest spatial correlations.

\subsection{Large scale MIMO} \label{secLGMIMO}
 The original measurements were taken using $4\times 4$ system. However, by manipulating the measurements we are able to create much larger virtual MIMO systems. 
 
  The following steps are taken in order to create the large scale  channel array:
  
 \begin{enumerate}
  \item Channel measurements from the reference device is used to exclude the body shadowing effects. 
  
  \item The original channels are reversed, such that the mobile station becomes the transmitting device. The reason for that is that the transmitters of the original channel measurements are fixed on top of a building, while the receiver device moved.
  
  \item The first channel from each snapshot, from the walking measurements, was chosen to form each of the virtual array transmitters, resulting in a virtual array with $1024$ elements.
  
  \item To reduce the correlation between adjacent channels, every fourth element of this array was chosen, forming a maximum array size of $256$ antennas. These are equally spaced along a path of about $6$ m in length.
  
  \item The locations with good fitting to Rayleigh fading distributions were first chosen, and then those that showed the lowest variation in their Rayleigh fading statistics between each virtual spatial channel were selected. This is done to avoid the scenario where the user experienced significant channel shadowing along part of the walking measurement, as this would introduce a significant power imbalance in the virtual MIMO channel. 
  
\end{enumerate}

 The Rayleigh fading mean statistic of the normalised constructed virtual MIMO channel has an average of 0.70, and a variance of 0.16. 
 \vspace{-0.3cm}
\section{Analytical Modelling of SM--ABER over Correlated and Uncorrelated Channels} \label{ANA}
 The ABER for SM system can be approximated by using the union bound \cite{book:p0001}, which can be expressed as follows, 
\begin{equation}
 \underset{\text{SM}}{\text{ABER}} \leq \sum_{\ell_t,s_t}\sum_{\ell,s}\frac{N\left({\bf{x}}_{\ell_t ,s_t },{\bf{x}}_{\ell ,s }\right)}{m}\frac{\mathrm{E}_{\bf{H}}\left\lbrace \Pr\left({\bf{x}}_{\ell ,s } \neq {\bf{x}}_{\ell_t ,s_t } \right) \right\rbrace }{2^m} \label{Eq:ABER_1}
\end{equation}
where $N\left({\bf{x}}_{\ell_t ,s_t },{\bf{x}}_{\ell,s}\right)$ is the number of bits in error between ${\bf{x}}_{\ell_t ,s_t }$ and ${\bf{x}}_{\ell,s}$, $\mathrm{E}_{\mathbf{H}}\{\cdot\}$ is the expectation across $\mathbf{H}$ and $\Pr\left({\bf{x}}_{\ell,s}\neq {\bf{x}}_{\ell_t ,s_t }\right)$ is the conditional pairwise error probability (PEP) of deciding on ${\bf{x}}_{\ell ,s }$ given that ${\bf{x}}_{\ell_t ,s_t }$ is transmitted,
\vspace{0.3cm}
\begin{eqnarray}
 \Pr\left({\bf{x}}_{\ell,s}\neq {\bf{x}}_{\ell_t ,s_t }\right) &=& \Pr\left(\left\| \mathbf{{y}} - \mathbf{{H}}{\bf{x}}_{\ell_t ,s_t}\right\|^2 > \left\| \mathbf{{y}} - \mathbf{{H}}{\bf{x}}_{\ell,s}\right\|^2 \right) \nonumber \\
 &=& Q\left(\sqrt{\frac{\left\|\mathbf{{H}}\Psi\right\|^2}{2\sigma_n^2}}\right) \label{Eq:Q_1} \nonumber \\
&=&\frac{1}{\pi} \int_0^{\frac{\pi}{2}}\exp\left(-\frac{\left\|\mathbf{{H}}\Psi\right\|^2}{4\sigma_n^2\sin^2\theta}\right)d\theta \label{Eq:Sin1}
\end{eqnarray} 
where $\Psi =  \left({\bf{x}}_{\ell_t ,s_t }-{\bf{x}}_{\ell,s}\right) $, and the alternative integral expression of the $Q$-function is given in~\cite{book:sa0001}.

 Taking the expectation of \eqref{Eq:Sin1}, we have,
\begin{equation}
 \mathrm{E}_{\bf{H}}\left\lbrace \Pr\left({\bf{x}}_{\ell,s}\neq {\bf{x}}_{\ell_t ,s_t }\right) \right\rbrace = \frac{1}{\pi}\int_0^{\frac{\pi}{2}}\Phi\left(-\frac{1}{4\sigma_n^2\sin^2\theta}\right)d\theta \label{Eq:MGF1}
\end{equation}
where $\Phi\left(\cdot\right)$ is the moment-generation function (MGF) of the random variable $\left\|\mathbf{\bar{H}}\Psi\right\|^2$.

 From \cite{hsn0501}, and noting that in SM only one antenna is active at a time, the MGF in \eqref{Eq:MGF1} for quasi--static fading with spatial correlation is equal to,
\begin{equation}
 \Phi\left(s\right) = \prod_{j=1}^{N_r}\left(1-s\lambda_j\mu\right)^{-1}\label{Eq:MGF2}
\end{equation}
where $\lambda_j$ are the eigenvalues of $\mathbf{R}_{\text{Rx}}$ and $\mu=|s_t|^2 + |s|^2 - 2  \text{Re}\{s_ts^*\} \mathbf{R}_{\text{Tx}}(\ell_t,\ell) $ . 

  Substituting \eqref{Eq:MGF2} and \eqref{Eq:MGF1} in \eqref{Eq:ABER_1} and using the Chernoff bound, the ABER for SM over Rayleigh channels is,
  
  \begin{equation}
 \underset{\text{SM}}{\text{ABER}} \leq \frac{1}{2\pi}\sum_{\ell_t,s_t}\sum_{\ell,s}\prod_{j=1}^{N_r}\frac{N\left({\bf{x}}_{\ell_t ,s_t },{\bf{x}}_{\ell,s}\right)}{m} \frac{1}{2^m}\left(1+\frac{\lambda_j\mu}{4\sigma_n^2}\right)^{-1} \label{Eq:ABER}
\end{equation}

In Section \ref{results}, we show that the two bounds; for uncorrelated and correlated Rayleigh channels, i) are tight upper bounds for  SM, and ii) they validate the experimental results.

\section{Complexity Analysis} \label{comp}

In this section the receiver complexity for SM--ML is compared to the complexity of SMX--ML. The complexity is computed as the number of real multiplicative operations $(\times,\div)$ needed by each algorithm \cite{book:gl9601}.
 \vspace{0.4cm}
\begin{itemize}
 \item  SM--ML: The computational complexity of the SML--ML receiver in \eqref{Eq:ML} is equal to  

 \begin{equation}
   \mathcal{C}_{\text{SM--ML}} = 8N_r2^m  \label{Eq:CompSM}
 \end{equation}
 
  as the ML detector searches through the whole transmit and receive search spaces. Note, evaluating the Euclidean distance $\left({\left| {y_r  - h_{\ell,r} s} \right|^2 }\right)$ requires $2$ complex multiplications, where each complex multiplication requires $4$ real multiplications. 
 \item SMX--ML: The computational complexity of SMX--ML is equal to,
 
 \begin{equation}
  \mathcal{C}_{\text{SMX--ML}} = 4\left(N_t+1\right)N_r2^m \label{Eq:CompSMX} 
  \end{equation}
  
\noindent Note, $\left(\left|\mathbf{y}-\mathbf{Hx}\right|^2\right)$ in \eqref{Eq:MLMIMO} requires $\left(N_t+1\right)$ complex multiplications.
\end{itemize}

 From \eqref{Eq:CompSM} and \eqref{Eq:CompSMX}, the reduction of SM--ML receiver complexity relative to the complexity of the SMX--ML decoder for the same spectral efficiency is given by,
 \begin{equation}
  \mathcal{C}_{\text{rel}} = 100 \times \left( 1 - \frac{2}{N_t+1} \right) \label{Eq:RelComp}
 \end{equation}
 
From \eqref{Eq:CompSM}, the complexity of SM does not depend on the number of transmit antennas, and it is equal to the complexity of single--input multiple--output (SIMO) systems. Hence, the reduction in complexity offered by SM increases with the increase in the number of transmit antennas. However, the complexity of SMX increases linearly with the number of transmit antennas. 
 For example from \eqref{Eq:RelComp}, for $N_t=4$, SM offers a $60\%$ reduction in complexity, and as the number of transmit antennas increase the reduction increases. For $N_t=128$, SM offers $98\%$ reduction in complexity.
\begin{figure}[!t]
    \centering
      \includegraphics[scale=0.56]{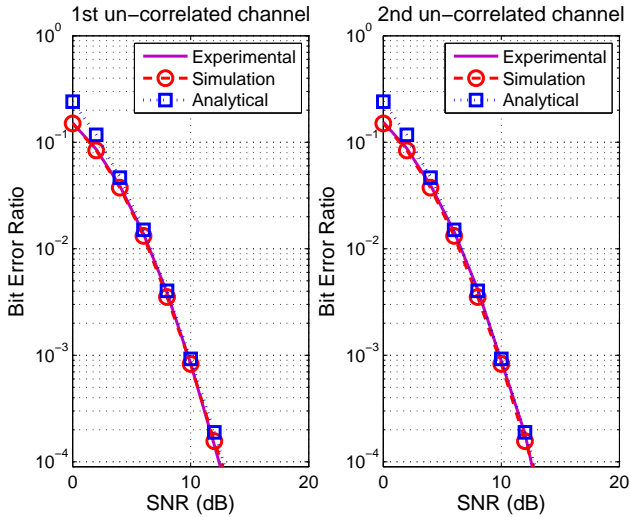}
      \caption{ ABER versus the SNR for SM over an uncorrelated channel. $m=4$, $N_t=N_r=4$.}
    \label{fig:BER_SM_uncorr}\vspace{-0.3cm}
 \end{figure}

 \begin{figure}[!t]
  \vspace{-0.12cm}
    \centering
      \includegraphics[scale=0.56]{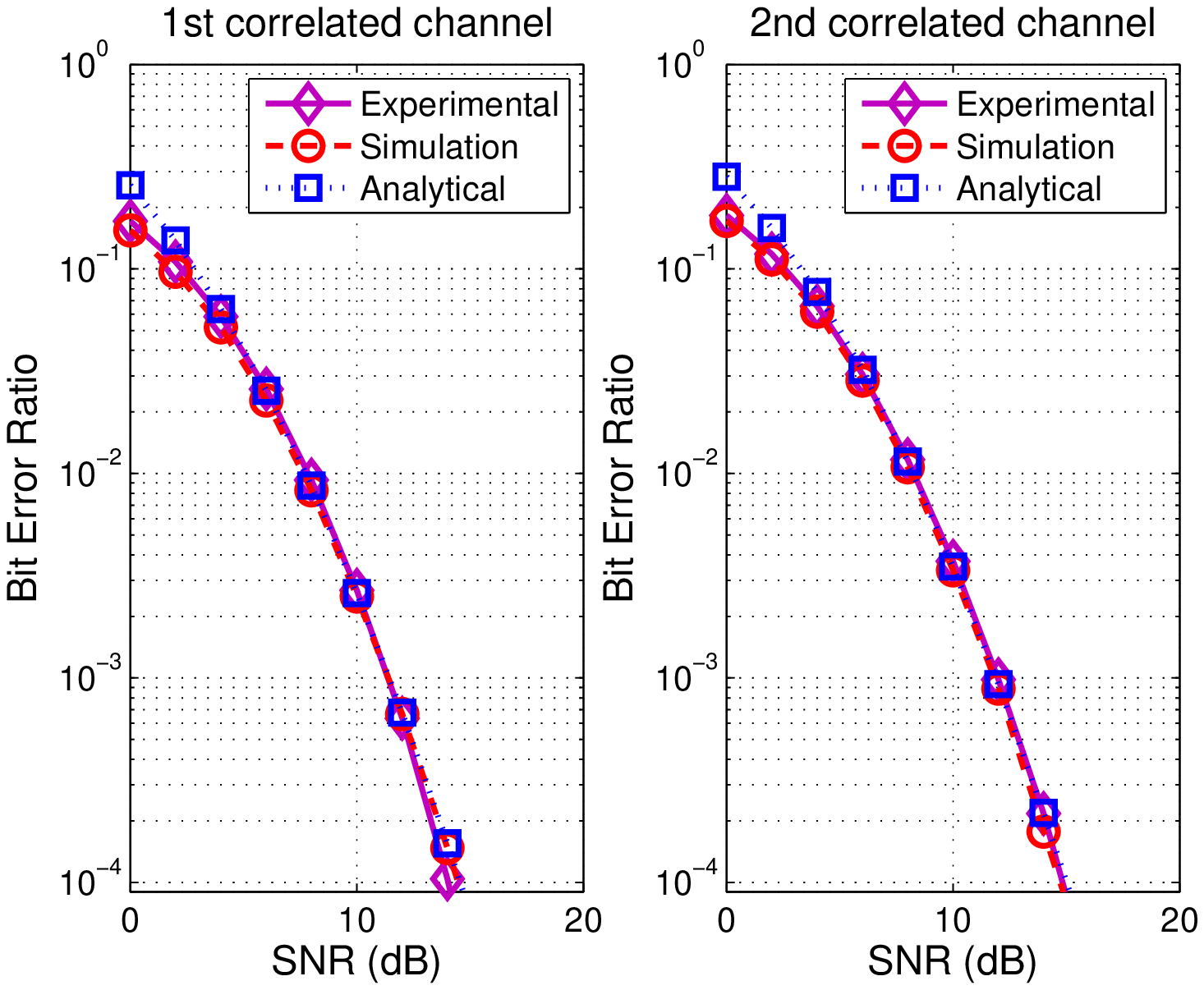}
      \caption{ ABER versus the SNR for SM over a correlated channel. $m=4$, $N_t=N_r=4$.}
    \label{fig:BER_SM_corr}
\end{figure}

\begin{figure}[ht]
    \centering
      \includegraphics[scale=0.56]{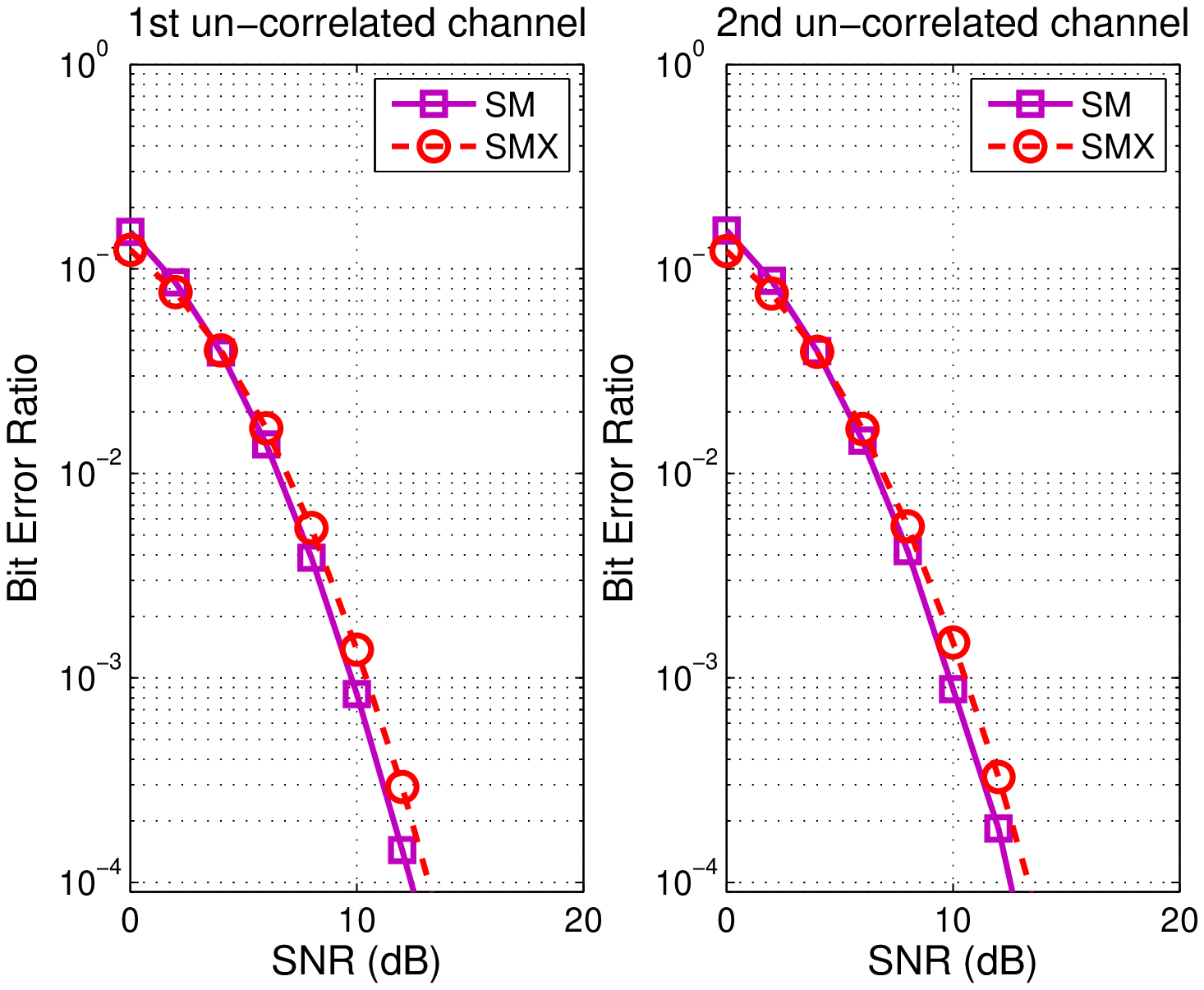}
      \caption{ ABER versus the SNR for SM and SMX over an uncorrelated channel. $m=4$, $N_t=N_r=4$.}
    \label{fig:BER_uncorr}
 \end{figure}

 \begin{figure}[ht]
    \centering
      \includegraphics[scale=0.56]{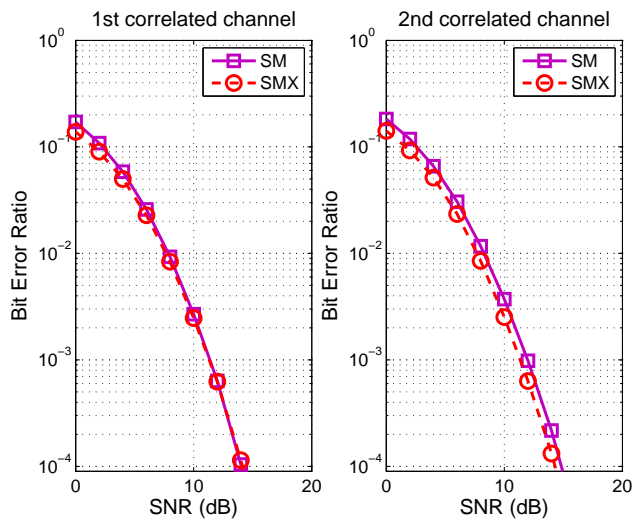} 
      \caption{ ABER versus the SNR for SM and SMX over a correlated channel. $m=4$, $N_t=N_r=4$.}
    \label{fig:BER_corr}
\end{figure}

\section{Results}\label{results}

In the following, Monte Carlo simulation results for the ABER performance of SM using the measured urban channels and simulated Rayleigh channels are compared with the derived analytical bound. Note, each channel of the measured urban channels contains $1024$ snapshots. Furthermore, the performance of SM using the measured urban channel is compared with the performance of SMX over the same channels for small and large scale MIMO.

\begin{figure}[!h]
    \centering
      \includegraphics[scale=0.46]{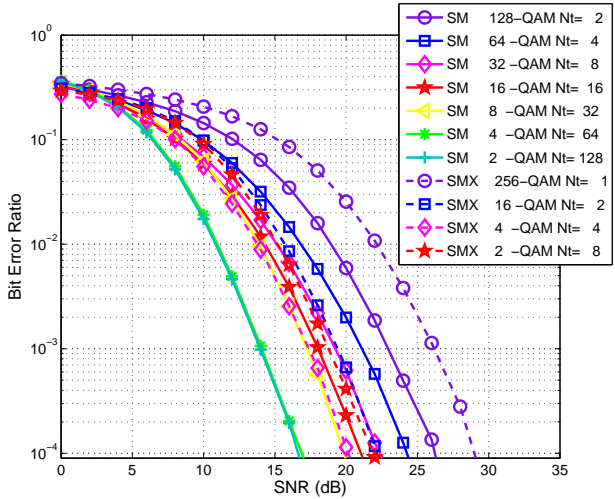} 
      \caption{ABER versus the SNR for SM and SMX over real measured channels. $m=8$, $N_r=4$.}     
    \label{fig:BER_All_Comb}\vspace{-0.6cm}
\end{figure}
\vspace{-0.2cm}

\subsection{Validation of SM analytical ABER performance using experimental results }

 Fig. \ref{fig:BER_SM_uncorr} and Fig. \ref{fig:BER_SM_corr} show the ABER performance of SM using the measured urban channels (solid line) and using simulated Rayleigh channels (red dashed line). The results are compared with the derived analytical bound (blue dotted line), for $m=4$ and $N_t=N_r=4$.
 Fig.~\ref{fig:BER_SM_uncorr} shows the ABER for uncorrelated channels and Fig.~\ref{fig:BER_SM_corr} shows the ABER for correlated channels. As can be seen from the figures, the experimental results closely match the simulation and analytical curves for ABER $<$ $10^{-2}$.
 In Fig. \ref{fig:BER_SM_uncorr} we can see that SM offers the same performance for both chosen channels, where both channels are uncorrelated. However, in Fig. \ref{fig:BER_SM_corr}, there is a slight difference in the performance, since the two chosen correlated channels have different correlation matrices. Moreover, if we compare the results for uncorrelated channels in Fig. \ref{fig:BER_SM_uncorr} with those correlated channels in Fig. \ref{fig:BER_SM_corr}, we see that SM performs better when the channels are uncorrelated channels, as it is easier to distinguish the different channel paths.
\vspace{-0.2cm}

\subsection{Comparison in the ABER performance of SM and SMX}

\subsubsection{Small Scale MIMO}

 Figs. \ref{fig:BER_uncorr} and \ref{fig:BER_corr} compare the ABER between SM (solid line) and SMX (dashed line) using the measured urban channels for $m=4$ and $N_t=N_r=4$. From both figures, we can see that SM offers almost the same as or slightly better performance than SMX. In Fig. \ref{fig:BER_uncorr}, the performance of both systems does not change for both channels since the channels are uncorrelated. 
 However, as shown in Fig.~\ref{fig:BER_corr}, this is not the case for the correlated channels, where the performance is different due to the different correlation coefficients.

\subsubsection{Large Scale MIMO}

 Fig. \ref{fig:BER_All_Comb} compares the ABER between SM (solid line) and SMX (dashed line) using the virtual large scale MIMO channel created using the measured urban channels as explained in Sec.\ref{secLGMIMO}, where $m=8$, $N_r=4$. For $m=8$ the maximum number of transmit antennas that SMX can use is $N_t=8$, where $m=N_t\log_2(M)$. However, for SM the maximum number of antennas that can be used is $N_t=128$, making it possible to exploit the advantages of large scale  MIMO. Note that for SM it holds that: $m=\log_2(N_t)+\log_2(M)$.
 Finally, in Fig. \ref{fig:BER_All_Comb} we can see that SM with $N_t=128$ and $N_t=64$ offers $6$~dB and $4$~dB better performance than SMX with $N_t=8$ and $N_t=4$ respectively. Note that the constellation size is the same for both SM with $N_t=128$ and SMX with $N_t=8$, as is for SM with $N_t=64$ and SMX with $N_t=4$. As the constellation size of the signal symbol is increased, the ABER of SM and SMX increases, {\it i.e.}, moving to $N_t=16$ for SM we see that SM offers only a $1$~dB performance increase relative to SMX with $N_t = 2$. Note, the number of bits sent per transmission for both SM and SMX for all the scenarios is equal,  $m=8$.

\section{Summary and Conclusion}\label{Con}
 In this paper, performance analysis of SM using urban Rayleigh channel measurements for both correlated and uncorrelated scenarios has been carried out. An analytical bound has been derived and performance results using simulated channels have been provided. An important observation is that experimental results confirm the analytical bound as well as computer simulations of the system. The performance of SM has been compared with the performance of SMX using the same urban channels. It has been demonstrated that for small scale MIMO, SM offers similar or slightly better ABER performance. However, for large scale  MIMO, SM exhibits a significant enhancement in the ABER performance at no increase in complexity. This makes SM an ideal candidate for future large scale MIMO systems.

\section*{Acknowledgement}
This work has been partially funded by the University of Edinburgh
Initiating Knowledge Transfer Fund (IKTF), RCUK (EP/G042713/1, UK-China Science Bridges ``R\&D on (B)4G
Wireless Mobile Communications") and the European Union
(PITNGA2010264759) ``GREENET" project.

 \bibliographystyle{IEEEtran}
%    \bibliography{}
 \bibliography{IEEEabrv,general,cwc}

\begin{thebibliography}{10}
\providecommand{\url}[1]{#1}
\csname url@samestyle\endcsname
\providecommand{\newblock}{\relax}
\providecommand{\bibinfo}[2]{#2}
\providecommand{\BIBentrySTDinterwordspacing}{\spaceskip=0pt\relax}
\providecommand{\BIBentryALTinterwordstretchfactor}{4}
\providecommand{\BIBentryALTinterwordspacing}{\spaceskip=\fontdimen2\font plus
\BIBentryALTinterwordstretchfactor\fontdimen3\font minus
  \fontdimen4\font\relax}
\providecommand{\BIBforeignlanguage}[2]{{%
\expandafter\ifx\csname l@#1\endcsname\relax
\typeout{** WARNING: IEEEtran.bst: No hyphenation pattern has been}%
\typeout{** loaded for the language `#1'. Using the pattern for}%
\typeout{** the default language instead.}%
\else
\language=\csname l@#1\endcsname
\fi
#2}}
\providecommand{\BIBdecl}{\relax}
\BIBdecl

\bibitem{t9902}
E.~Telatar, ``{Capacity of Multi-Antenna Gaussian Channels},'' \emph{European
  Trans. on Telecommun.}, vol.~10, no.~6, pp. 585--595, Nov. / Dec. 1999.

\bibitem{f9601}
G.~J. Foschini, ``{Layered Space-Time Architecture for Wireless Communication
  in a Fading Environment when Using Multi-Element Antennas},'' \emph{{Bell
  Labs Tech. J.}}, vol.~1, no.~2, pp. 41--59, 1996.

\bibitem{cv0901}
B.~Cerato and E.~Viterbo, ``{Hardware implementation of a low-complexity
  detector for large MIMO},'' in \emph{{IEEE Int. Symp. on Circuits and Systems
  (ISCAS 2009)}}, May 2009, pp. 593 --596.

\bibitem{mhsay0801}
R.~Mesleh, H.~Haas, S.~Sinanovi\'{c}, C.~W. Ahn, and S.~Yun, ``{Spatial
  Modulation},'' \emph{{IEEE Trans. on Veh. Tech.}}, vol.~57, no.~4, pp. 2228
  -- 2241, Jul. 2008.

\bibitem{jgsc0901}
J.~Jeganathan, A.~Ghrayeb, L.~Szczecinski, and A.~Ceron, ``{Space Shift Keying
  Modulation for MIMO Channels},'' \emph{{IEEE Trans. on Wireless Commun.}},
  vol.~8, no.~7, pp. 3692--3703, Jul. 2009.

\bibitem{kspmf0201}
J.~P. Kermoal, L.~Schumacher, K.~I. Pedersen, P.~E. Mogensen, and
  F.~Frederiksen, ``{A Stochastic MIMO Radio Channel Model with Experimental
  Validation},'' \emph{IEEE Journal on Selected Areas in Communications},
  vol.~20, no.~6, pp. 1211 -- 1226, Aug. 2002.

\bibitem{vzh0201}
A.~V. Zelst and J.~S. Hammerschmidt, ``{A Single Coefficient Spatial
  Correlation Model for Multiple-Input Multiple-Output (MIMO) Radio
  Channels},'' in \emph{27th General Assembly of the International Union of
  Radio Science {(URSI)}}, 2002, pp. 1--4.

\bibitem{hb0601}
M.~Hunukumbure and M.~Beach, ``{MIMO Channel Measurements and Analysis with
  Prototype User Devices in a 2GHz Urban Cell},'' in \emph{2006 IEEE 17th
  International Symposium on Personal, Indoor and Mobile Radio Communications
  (PIMRC)}, Sept. 2006, pp. 1 --5.

\bibitem{l1201}
{INOTEK Antennas}, ``{Single Band, Dual Polarised Antenna {UMTS}
  {XP}/65/17.5/2, 4, 6, 8 or 10 Type 2209},''
  http://www.inotekantennas.com/pdf/2209.pdf.

\bibitem{book:p0001}
J.~G. Proakis, \emph{{Digital Communications}}, 4th~ed.\hskip 1em plus 0.5em
  minus 0.4em\relax {McGraw--Hill}, 2000.

\bibitem{book:sa0001}
M.~K. Simon and M.-S. Alouini, \emph{{Digital Communication over Fading
  Channels: A Unified Approach to Performance Analysis}}, 1st~ed.\hskip 1em
  plus 0.5em minus 0.4em\relax John Wiley \& Sons, Inc., 2000.

\bibitem{hsn0501}
A.~Hedayat, H.~Shah, and A.~Nosratinia, ``Analysis of space-time coding in
  correlated fading channels,'' \emph{IEEE Transactions on Wireless
  Communications}, vol.~4, no.~6, pp. 2882 -- 2891, Nov. 2005.

\bibitem{book:gl9601}
G.~H. Golub and C.~F. van Loan, \emph{{Matrix Computations}}.\hskip 1em plus
  0.5em minus 0.4em\relax {The John Hopkins University Press}, 1996.

\end{thebibliography}

\end{document}